\documentclass[sigconf]{acmart}

\usepackage{tikz}
\usepackage{amsmath}
\usepackage{tabularx,xcolor,colortbl}
\usepackage{multirow}
 
\usepackage{listings}
\usepackage{paralist}
\usepackage{enumitem}

\usetikzlibrary{shapes,arrows}
\usepackage{algorithm}
\usepackage{multirow}
\usepackage{booktabs}
\usetikzlibrary{shapes, arrows.meta}
\usepackage{tabularx}
\usetikzlibrary{positioning}
\usepackage[section]{placeins}
\usepackage{fontawesome} 
\usepackage{algorithmic}
\usepackage{xcolor}
\usepackage{multicol}
\usepackage{tcolorbox}
\usepackage{listingsutf8}

\usepackage{graphicx}

\usepackage[compact]{titlesec}
\titlespacing{\section}{0pt}{2ex}{1ex}

\setcopyright{acmcopyright}
\copyrightyear{2024}

\acmYear{YYYY}
\acmVolume{YYYY}
\acmNumber{X}
\acmDOI{XXXX.XXXX}
\acmISBN{}
\acmConference{LADC 2024}

\begin{document}

\title[Feature Selection for Web Applications Attack Detection]{Capturing the security expert knowledge in feature selection for web application attack detection\\ \large{Research paper}}


\author{Amanda Riverol}
\affiliation{%
  \institution{Facultad de Ingeniería, Universidad de la República}
  \city{Montevideo}
  \country{Uruguay}}
\email{ariverol@fing.edu.uy}

\author{Gustavo Betarte}
\affiliation{%
  \institution{Facultad de Ingeniería, Universidad de la República}
  \city{Montevideo}
  \country{Uruguay}}
\email{gustun@fing.edu.uy}

\author{Rodrigo Martínez}
\affiliation{%
  \institution{Facultad de Ingeniería, Universidad de la República}
  \city{Montevideo}
  \country{Uruguay}}
\email{rodmart@fing.edu.uy}

\author{Álvaro Pardo}
\affiliation{%
  \institution{Departamento de Ingeniería,  Universidad Católica del Uruguay}
  \city{Montevideo}
  \country{Uruguay}}
\email{apardo@ucu.edu.uy}


\renewcommand{\shortauthors}{Riverol et al.}

\begin{abstract}
This article puts forward the use of mutual information values to replicate the expertise of security professionals in selecting features for detecting web attacks. The goal is to enhance the effectiveness of web application firewalls (WAFs). Web applications are frequently vulnerable to various security threats, making WAFs essential for their protection. WAFs analyze HTTP traffic using rule-based approaches to identify known attack patterns and to detect and block potential malicious requests. However, a major challenge is the occurrence of false positives, which can lead to blocking legitimate traffic and impact the normal functioning of the application. The problem is addressed as an approach that combines supervised learning for feature selection with a semi-supervised learning scenario for training a One-Class SVM model. The experimental findings show that the model trained with features selected by the proposed algorithm outperformed the expert-based selection approach in terms of performance. Additionally, the results obtained by the traditional rule-based WAF ModSecurity, configured with a vanilla set of OWASP CRS rules, were also improved.
\end{abstract}

\keywords{Web applications, Attack Detection, Models,Feature Selection}

\maketitle


\section{Introduction}
\label{sec:intro}
A web application operates within a client-server architecture, where the server handles computational tasks like data transmission, processing, and storage, while the client interacts via a web browser. These applications face significant security risks \cite{romanartificial}. Vulnerabilities, spanning from design through implementation and configuration, pose threats to data integrity, confidentiality, and availability.

To address these concerns, the Open Web Application Security Project (OWASP) \cite{owasp} maintains the OWASP Top Ten, listing the most critical security risks to web applications. Web Application Firewalls (WAFs), as defined by Ghanbari \cite{ghanbari2015comparative}, act as security checkpoints, analyzing and blocking HTTP traffic to identify potential malicious requests. ModSecurity, a widely used open-source WAF, relies on the Core Rule Set (CRS) compiled by OWASP to detect known attack patterns. The CRS, recognized as a standard in the industry, includes rules crafted by experts to detect variants of attacks with different levels of  severity \cite{montaruli2023adversarial}.

WAF solutions like ModSecurity assess HTTP requests by computing an overall score based on activated rules. However, false positives remain a challenge, potentially disrupting legitimate traffic and normal application functions. Addressing this issue involves configuring the CRS, a task that can be complex for non-security professionals. Recent research \cite{betarte2018web, rmartinez-ladc-2018, montes2021web}  indicates that machine learning models can enhance attack detection, often outperforming traditional methods such as rule-based static analysis and signature-based attack patterns without requiring extensive security expertise \cite{montes2021web}.

\begin{figure}[h!]
    \centering
    \includegraphics[width=0.45\textwidth]{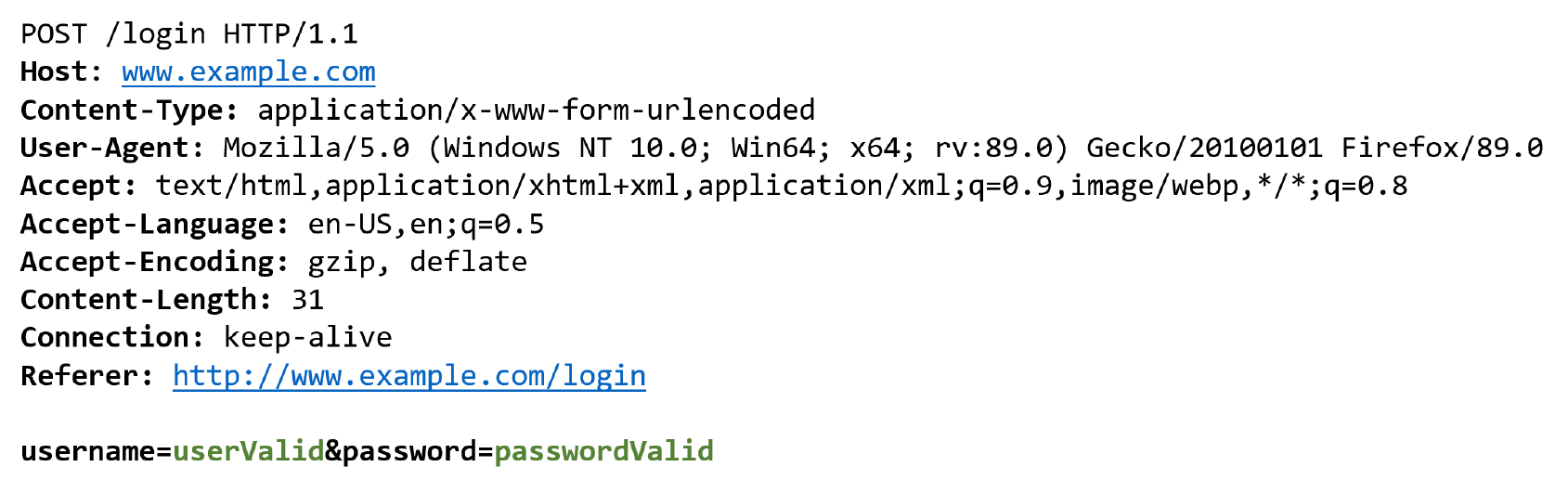}
    \caption{Valid Request}
    \label{fig:valid}
\end{figure}

\begin{figure}[h!]
    \centering
    \includegraphics[width=0.45\textwidth]{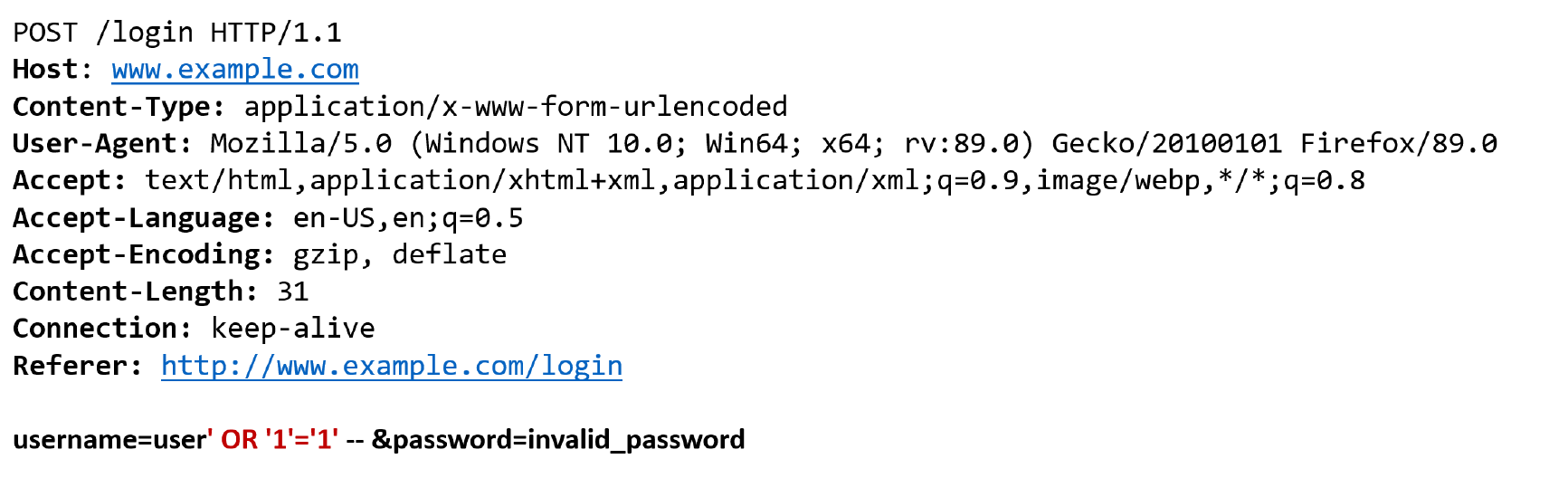}
    \caption{Attack Request (SQL Injection)}
    \label{fig:attack}
\end{figure}

The biggest problem when trying to differentiate valid from anomalous requests is their similarity.  Consider the following example: distinguishing between a valid request (Figure \ref{fig:valid}) and an attack request, such as a SQL injection attempt (Figure \ref{fig:attack}).
While both requests may initially appear similar in terms of standard HTTP headers and parameters, the fundamental distinction lies in specific tokens within the query. For example, the presence of the token \textbf{OR '1'='1'} in a SQL injection request (Figure \ref{fig:attack}) denotes an attempt to bypass authentication by injecting malicious SQL code, a clear indicator of an attack. In contrast, valid request typically do not feature these types of constructions.

This differentiation underscores the importance of focusing efforts to detect these anomalies on specific tokens. In this way, by performing automated learning focused on prioritizing these critical tokens, security systems could significantly improve the detection of anomalous requests and will reduce the noise that similar tokens can introduce between them and a valid request. This detection strategy is focused on not only improving the ability to detect new attack variants, but also mitigating the impact of false positives, thus protecting web applications more effectively against evolving security threats.

This article discusses how to enhance anomaly detection in web applications by automatcally selecting specific features. We focus on a methodology based on mutual information to identify and prioritize critical features (tokens) that differentiate between benign and malicious requests. While security experts traditionally make this selection, relying solely on experience or human intuition can introduce subjectivity and overlook emerging attack vectors or subtle patterns. Therefore, our approach advocates for a data-driven selection process that uses machine learning to objectively identify and prioritize features with high discriminatory power. Our research aims to improve the accuracy and effectiveness of anomaly detection systems in web applications by identifying and prioritizing critical features. This work contributes by providing a reproducible methodology to enhance the security of various web applications.

Two contributions are presented to the field of web application security. These contributions strengthen web application security against emerging threats and enhance the performance of WAFs:
\begin{itemize}
    \item First, the use of a dataset with diverse attack types is introduced, allowing feature selection without relying on application-specific data that is often difficult to obtain or requires expert labeling.
    \item We introduce a method for selecting features based on mutual information values, executed using only normal application traffic and the aforementioned attack set. The selection obtained by this method is used to train a one-class model, which improves its performance in anomaly detection. This improvement is achieved by more effectively identifying unique features between benign and malicious traffic compared to those chosen by experts.
\end{itemize}

The structure of the remainder of the article is as follows: Section \ref{sec:backg} addresses vulnerabilities in web applications and the use of machine learning or deep learning to enhance WAF performance by reducing false positives. Each stage of the learning process is described, emphasizing how feature selection can optimize its effectiveness. Section \ref{sec:implement} focuses on feature selection using mutual information. The results obtained are discussed in Section \ref{sec:discuss}, followed by an analysis of related work in Section \ref{sec:relwork}. Finally, Section \ref{sec:concl} presents the study's conclusions and outlines future research directions.

\section{Background}
\label{sec:backg}
Below is a brief overview of web applications and their security issues. We will examine attack detection models in web applications for protection and the issues and limitations they face in terms of generalization and adaptability. Additionally, we will delve into various learning techniques used in anomaly detection to gain a better understanding of their effectiveness and applicability in web security environments. This analysis will form the basis for exploring the techniques used in the later stages of the research.

\subsection{Web Applications and Vulnerabilities}
Web applications are fundamental in our lives, used in organizations and daily activities, handling large amounts of data and personal information. This type of information is crucial for the internal functioning of organizations and is of interest to third parties and even governments. However, accessing this information often comes at a high cost, leading some individuals to resort to illegal means, such as attacks on web applications, to obtain the desired information. These attacks can have various intentions, from extortion, fraud, and identity theft to manipulating the web application's reputation.

The lack of security in web applications constantly exposes them to risks as attackers exploit vulnerabilities in their infrastructure. One of the leading causes is the absence of security properties, such as logical correctness, input validity, state integrity, or adequate security configuration. For example, the lack of input validation and sanitization can introduce untrusted special characters, leading to common attacks like SQL injection, cross-site scripting (XSS) vulnerabilities, and cross-site request forgery (CSRF) attacks. This topic is addressed in a survey conducted on \cite{chaudhari2014survey}, which provides a state-of-the-art web application security analysis focusing on the challenges of creating secure web applications. This study highlights the importance of addressing existing security vulnerabilities to ensure adequate protection of web applications.

Therefore, it is essential to implement effective security measures, such as proper input validation and data sanitization, along with the use of tools like Web Application Firewalls (WAFs), which, as previously mentioned, act as a security checkpoint between users and the web application. Traditional WAFs use rule-based approaches to identify known attack patterns. However, these systems have significant limitations, especially against zero-day attacks, where existing rules cannot recognize new threats. Moreover, configuring and maintaining these systems can be complex and prone to errors, require the intervention of a security expert, and can result in high rates of false positives and negatives.

The evolution of AI techniques has drawn attention to the application of machine learning to anomaly detection through WAFs. In an analysis conducted by Applebaum et al. \cite{applebaum2021signature}, the issues with signature-based systems are highlighted, such as their vulnerability to zero-day attacks and the complexity of their configuration. The effectiveness of machine learning-based WAFs as an alternative is emphasized, and areas for improvement have been noted, such as performance evaluation on current hardware and application to different types of attacks.

\subsection{Automated learning for improving WAF performance}
Recent research \cite{betarte2018web, rmartinez:thesis, montes2021web} indicates that the detection of attacks using machine learning models reduces false positives when compared with the detection performed by the ModSecurity WAF configured with the CRS as a baseline. The model discussed in \cite{montes2021web} overcomes these results without requiring extensive security experience using a one-class approach combined with an automatic estimation of the best operational point. 

To build a machine learning model for attack detection, there are two alternatives.
The \textbf{multiclass approach} assumes that you have valid and attack requests for the application. We have implemented this approach using several classifiers for attack detection and including a preprocessing stage that uses knowledge of the HTTP structure to improve feature extraction ~\cite{betarte2018improving,rmartinez:thesis}. Our experiments have validated the effectiveness of this approach. However, according to our extensive study, training the model with generic data sets and testing it with application-specific data, has revealed that classifiers built this way do not generalize well. This means that a model trained for one application cannot be directly applied to protect a different one.

In situations where there are only available requests that belong to the valid or attack class, we have explored a \textbf{one-class classification approach}. This approach is discussed in ~\cite{montes2021web} in which requests are analyzed by counting the occurrence of specific attributes. These attributes, which best define the different attacks on web applications, were determined with the input of a security expert. A key aspect of this approach is the threshold that adjusts the classification into valid or attack. Each potential threshold value represents an operating point of the model, allowing the expert to modify the attack detection or false positive rate simply by altering this value.

Regarding the one-class approach model, we have been exploring alternatives to automatically select the optimal operating point using sampling or synthetic attacks. We have experimented with algorithms of a class like SVM and deep learning techniques to extract the features. The first results of this line of work have been presented in~\cite{montes2021web}.

The article \cite{gniewkowski2023sec2vec} delves into advances in anomaly detection using natural language processing techniques. It underlines the need for an improved tokenizer capable of handling tokens beyond the standard vocabulary, crucial for detecting emerging attack patterns. While most data sets showed consistent results, there was a notable decrease in the performance of the RoBERTa model, highlighting a challenge associated with changing concepts. It also shows the effectiveness of Bag-of-Words (BoW) in identifying abnormal segments within web request data, particularly with the integration of percentage-encoded characters into an optimized dictionary, thus improving the accuracy of local anomaly detection. Furthermore, the importance of custom tokenization is emphasized to reinforce the adaptability and effectiveness of BoW in real-world cybersecurity scenarios.

\subsection{Preprocessing and Tokenization}
In the anomaly detection process, each stage can vary depending on the specific approach applied, and errors or deficiencies in any of these stages can significantly affect the performance of the resulting models \cite{salehin2024automl}.

The tokenization stage is what allows a text to be divided into smaller parts called tokens. These tokens are later used to find patterns and are considered a basic step in stemming. Tokenization also helps replace sensitive data elements with non-sensitive data elements.

Below we will explore the most common techniques used for representating or vectorizing of these tokens, which serve as input for machine learning models.

Then, we will emphasize the next stage of feature selection and its contribution to anomaly detection. We focus our work with the hypothesis that the tokens selected in this stage influence the final results of the methods.

\subsubsection{Vectorization methods} Vectorization methods are techniques used to convert textual or categorical data into a numerical representation that machine learning models can process. Below are some of the most common vectorization methods used in machine learning-based WAF implementations:

\textbf{BoW}. BoW (Bag-of-words) is a simple model that represents a document using the frequency of words (every position in the vector corresponds to a word) or, in our example, tokens obtained from the tokenization stage. The vector size is limited by using only the most common tokens. Ren et al. \cite{ren2018web} demonstrate the effectiveness of BoW in extracting features for web attack detection using hidden Markov algorithms. Their research shows improved detection rates and reduced false positives compared to previous experiments utilizing N-grams.
Mathematically, the vectorization using BoW can be expressed as:
\[
\mathbf{x}_i = \left[ x_{i1}, x_{i2}, \ldots, x_{iM} \right]
\]
where
\[
x_{ij} = \text{count}(w_j,d_i)
\]
Here, $\text{count}(w_j, d_i)$ denotes the number of occurrences of the word $w_j$ in the document $d_i$.

The Bag-of-Words (BoW) model simplifies text analysis by representing documents as word frequency vectors. It's versatile, interpretable, and efficient for various Natural Language Processing (NLP) tasks. However, it loses context, creates high-dimensional featue vectors, suffers from sparsity, and lacks semantic understanding.

\textbf{TF-IDF:} (Term Frequency-Inverse Document Frequency) stands as another common technique for feature extraction in cyber threat detection. Unlike BoW, TF-IDF counts the frequency of occurrence of each word in a document and weights these frequencies based on the word's importance in the entire document corpus.
Mathematically, the vectorization using TF-IDF can be expressed as:
\[
\text{TF-IDF}(w_j, d_i) = \text{TF}(w_j, d_i) \times \text{IDF}(w_j)
\]
Where:

\begin{itemize}
    \item \( \text{TF}(w_j, d_i) \) represents the term frequency of word \( w_j \) in document \( d_i \), calculated as the ratio of the number of occurrences of word \( w_j \) to the total number of words in document \( d_i \).
    \item \( \text{IDF}(w_j) \) denotes the inverse document frequency of word \( w_j \), calculated as the logarithm of the ratio between the total number of documents in the corpus and the number of documents containing word \( w_j \).
\end{itemize}

TF-IDF improves upon BoW by weighting word frequencies based on their importance in the corpus. It is beneficial for capturing the relevance of terms within documents and across the corpus, aiding in feature interpretation and reducing the impact of common terms. However, TF-IDF can still suffer from high dimensionality and sparsity, especially in large datasets, and requires careful tuning to achieve optimal performance.

\textbf{Word2Vec:} is a deep neural network-based technique that maps vectors of real numbers to words in a low-dimensional vector space. Word2Vec captures the semantic and syntactic relationships between words and represents similar words with vectors close to each other in the vector space. An example of using these techniques to transform the content of web pages into word vectors, allowing the representation of words in a continuous vector space, can be seen in the works \cite{li2020weighted, liu2021http}.

\textbf{RoBERTa:} is a transformer-based language model \cite{liu2019roberta} that builds upon the BERT architecture \cite{devlin2018bert}. Unlike BERT, RoBERTa removes the next-sentence prediction (NSP) objective and introduces several enhancements. These include dynamic masking, training on full sentences without NSP loss, using larger mini-batches, and employing a larger byte-level Byte Pair Encoding (BPE). Additionally, RoBERTa emphasizes the importance of the pretraining data and the number of training passes through the data. These modifications collectively improve token representation and end-task performance in documents without clear sentence boundaries.

These are the four main text representation techniques in natural language processing (NLP). Bag-of-Words and TF-IDF are simple yet effective methods for feature extraction, whereas Word2Vec and RoBERTa offer more advanced representations by capturing semantic and contextual relationships in the text. Although Bag-of-Words and TF-IDF are more straightforward to understand and apply, they cannot model semantics and context. Conversely, Word2Vec and RoBERTa can capture these linguistic complexities but require more computational resources and may face challenges with rare words or domain-specific contexts.

Since this study focuses on feature selection, we will evaluate its effectiveness through a one-class model trained to classify normal web application traffic. The selected features will be vectorized using the Bag-of-Words technique to demonstrate that even with a simple vectorization, the feature selection performs comparably to a model trained with features chosen by security experts.

\subsection{Feature Selection} 
Feature selection, as a dimensionality reduction technique, aims to choose a small subset of relevant features from the original features by removing irrelevant, redundant, or noisy features. Feature selection can lead to higher learning performance, lower computational cost, and better model interpretability. Recently, researchers in computer vision, text mining, etc., have proposed a variety of feature selection algorithms and shown the effectiveness of their works in terms of theory and experiment.

In a review of the state of the art on these techniques \cite{miao2016survey}, a comprehensive experiment is conducted to test whether feature selection can improve learning performance by showing that feature selection benefits machine learning tasks. Feature selection methods are usually classified into three main types: filter, envelope, and embedded.

\begin{itemize}
    \item {\bf Filter methods} evaluate features independently of the learning model, using statistical measures such as Pearson correlation, Linear Discriminant Analysis (LDA), ANOVA, Chi-square test, Wilcoxon Mann Whitney test, and Mutual Information. Mutual Information measures the dependency between two variables and selects features with the highest dependency on the target variable \cite{cover1991elements}. These techniques reduce dimensionality by selecting features based on their relationship with the response variable before applying any learning algorithm.
    \item{\bf Wrapper methods} consider the interaction between features and the learning algorithm. These methods evaluate subsets of features by building and assessing a model. Although potentially more accurate, wrapper methods are computationally intensive. Examples include Recursive Feature Elimination (RFE) \cite{guyon2002gene} and forward selection algorithms \cite{kohavi1997wrappers}.
    \item{\bf Embedded Methods} perform feature selection during the model training process. Examples include decision trees \cite{breiman1984classification} and regularization methods like Lasso \cite{tibshirani1996regression}, which penalize model complexity by including only significant features.
\end{itemize}

Among the various feature selection methods available, we have specifically chosen mutual information because it can measure the dependence between features and the target variable (class). Mutual information is a powerful statistical measure that evaluates the relevance of features based on their relationship with the target variable before applying any learning algorithm.

\textbf{Entropy and Mutual Information}:  Mutual information is an effective statistical tool for performing feature selection using filtering methods \cite{beraha2019feature}. In this context we will introduce entropy and mutual information.

The entropy \( H(X) \) of a random variable \( X \), with probability density function \( p \), measures uncertainty:
\[
H(X) := \mathbb{E}_{X} [-\log p(X)] = - \int p(x) \log p(x) \, dx.
\]
The integral 
calculates the expected value of the quantity \(-\log p(x)\), which represents the "self-information" associated with each value of \(X\). The result is the mean information of \( X \), or in other words, a global measure of the uncertainty in \(X\).

The mutual information \( I(X; Y) \) between two random variables \( X \) and \( Y \) is defined as:

\[
\begin{aligned}
I(X; Y) & := H(Y) - H(Y | X) \\
        & = \mathbb{E}_{X} \left[ D_{\text{KL}}(p(Y | X) \| p(Y)) \right] \\
        & = \iint p(x, y) \log \frac{p(x, y)}{p(x) p(y)} \, dx \, dy.
\end{aligned}
\]
The double integral calculates the expected value of the quantity \(\log \frac{p(x, y)}{p(x) p(y)}\), which represents how much the joint distribution \(p(x, y)\) differs from the distribution of \(X\) and \(Y\) if they were independent (\(p(x) p(y)\)). The result is a measure of the dependence between \(X\) and \(Y\).

Intuitively, the Mutual Information (MI) between $X$ and $Y$ represents the reduction in the uncertainty of $Y$ after observing $X$ (and vice versa). Notice that the MI is symmetric, i.e., $I(X; Y) = I(Y; X)$.

\section{Feature Selection using Mutual Information}
\label{sec:implement}
In previous studies \cite{rmartinez:thesis}, supervised machine learning models were implemented that required the intervention of a security expert to select features relevant to attack detection. Although these studies demonstrated good results, the need for a labeled set of valid traffic and attacks makes their application in real environments difficult.
As an alternative, a supervised model of one class \cite{montes2021web} was implemented that combined RoBERTa as a feature extractor and One-Class SVM. This approach managed to reduce false positives and demonstrated good performance, as well as eliminating the dependency on application-specific attack sets and the need for experts for feature selection. However, once trained cannot be reused in other applications and the training stage has a high computational cost.

The implementation we present in this section proposes to train a semi-supervised model of a One-Class SVM using Bag of Words as an extraction method, and incorporate a feature selection stage based on mutual information values.

To allow mutual information to capture distinctive features present in attacks, distinguishing between a valid request and an attack, we use a dataset with several types of attacks. This data set is intended to increase the likelihood that the algorithm will value tokens associated with attacks and not just limit itself to tokens present in valid requests. The objective is to demonstrate that these attacks do not necessarily have to be specific to the application; they can be generic or evolve over time, incorporating attacks from various applications, and still yield good results. This approach is feasible because constructing this dataset is more practical than obtaining specifically labeled attacks for the application being protected.
\subsection{Datasets}
To implement our proposed methodology for feature selection in web attack detection, we have developed a dataset of diverse types of attack. Additionally, to complement this attack dataset, it is necessary to generate a set of normal requests from the target application. These requests will represent the typical traffic that the application experiences during legitimate use. The inclusion of this normal data is crucial as it provides a clear contrast with attacks, enabling the model to distinguish between benign behaviors and suspicious activities.

Our proposed methodology relies on the combination of these two datasets: the diverse attack dataset and the set of normal application requests. By utilizing both datasets, we can select features relevant to both attacks and normal application operations.

\subsubsection{Attack Datasets}
\label{subsubsection:attackdatasets}
Creating a dataset of attacks for a specific application is a complex process involving the collection and labeling of representative data from various types of web attacks. Samples of network traffic containing malicious activities are selected for this purpose. This dataset should include a wide variety of attacks such as SQL injections, cross-site scripting (XSS) and Command Injection . Proper construction of this dataset requires the involvement of cybersecurity experts to accurately label the data, ensuring each instance is correctly classified.

Adding complexity to the process, it is essential that the distribution of attacks in the dataset is balanced to ensure detection models are not biased towards a specific type of attack. A proper representation of each type of attack is crucial for generalizing the model to real-world scenarios.

The goal of constructing a generic attack dataset lies in the need for a dataset that can be used in feature selection processes, regardless of the specific application, combined with valid traffic from the application itself. The advantages of a generic dataset include its applicability across multiple contexts and the ability to compare different detection methods under similar conditions.

\subsubsection{Classification and Distribution of Attacks}
\begin{figure}[h!]
    \centering
    \includegraphics[width=1\linewidth]{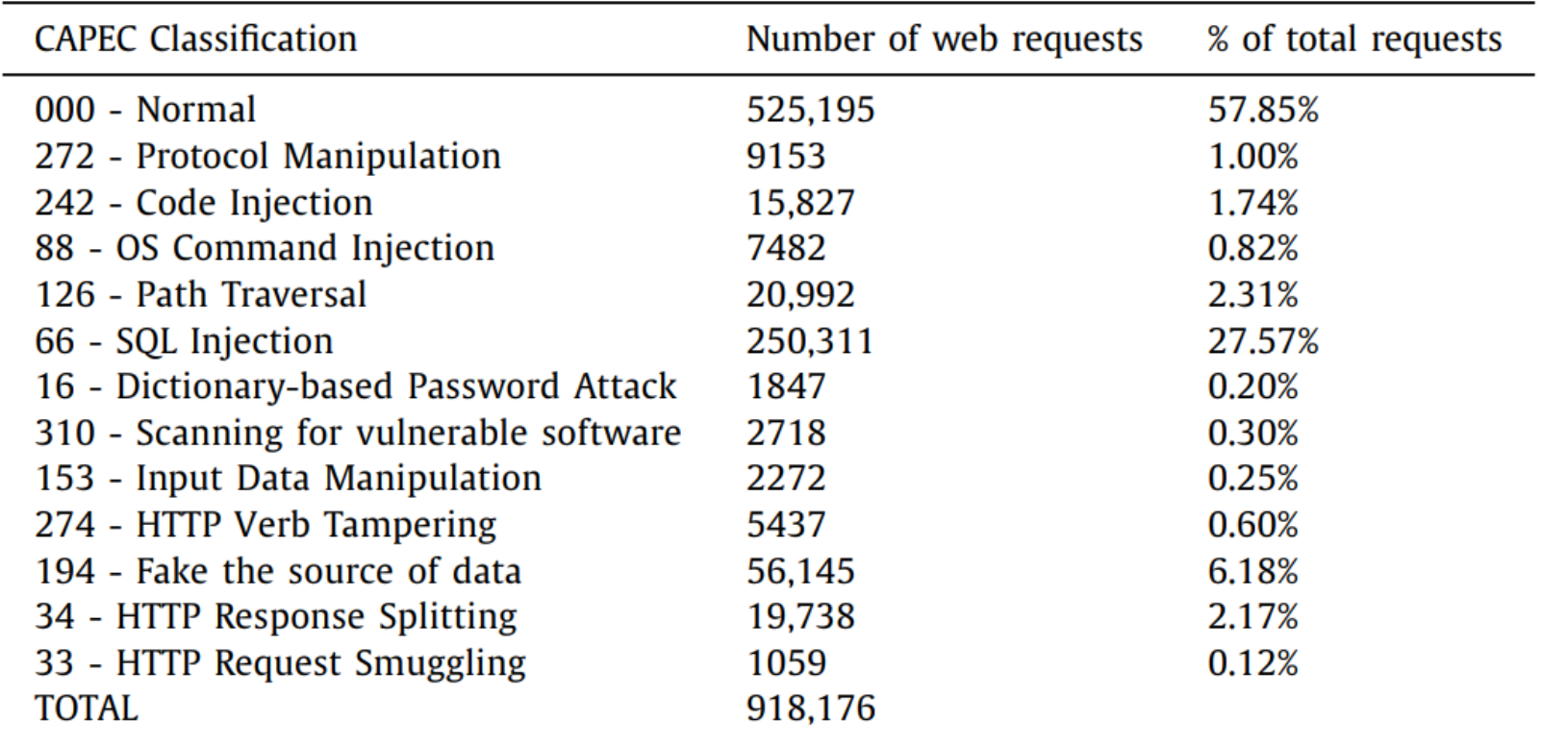}
    \caption{SR-BH 2020 Attack Distribution \cite{riera2022new}.}
    \label{fig:attack_distribution}
\end{figure}

The construction of the attack dataset used attacks present in the \textbf{SR-BH 2020} \cite{riera2022new} datasets, which include a wide variety of attacks classified as seen in the Figure \ref{fig:attack_distribution}.

The dataset is based on collecting real traffic in a honeypot exposed to the Internet for 12 days. This dataset includes a set of 13 different labels, which provide information about the normality of each web request and its possible classification into 12 different CAPEC (Common Attack Pattern Enumeration and Classification) categories. To create the set of attacks, 15\% of each category was taken.
 
\begin{figure}[h!]
    \centering
    \includegraphics[width=1\linewidth]{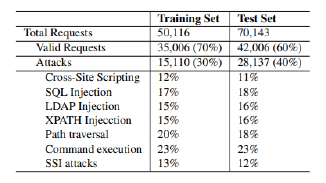}
    \caption{PKDD Attack Distribution  \cite{gallagher2009classification}.}
    \label{fig:attack_distribution_pkdd}
\end{figure}

Additionally, as a second dataset, we use the attacks from the PKDD dataset created for the PKDD2007 challenge \cite{ecml-pkdd-challenge}. This dataset includes real-world web traffic data collected during a specific period, labeled as normal and attacks, as presented in the study by Gallagher et al. \cite{gallagher2009classification}. PKDD offers a variety of cyber attacks, and the distribution of the dataset can be seen in Figure \ref{fig:attack_distribution_pkdd}.

The PKDD dataset contains less types of attacks compared to the SR-BH 2020 dataset. Although the data quality is lower, it offers the opportunity to investigate the use of a dataset with varied attack types.

The reuse of previously labeled attack data allows leveraging the prior work of data labeling and enrichment, significantly reducing the cost and time required for creating new datasets. Furthermore, data reuse facilitates the creation of representative and diverse datasets, enhancing the ability to generalize to new situations and detect a wide range of attacks.

\subsection{Preprocessing Stage}
\label{subsection:preprocessingstage}
This stage is aimed to enhance the quality and coherence of data, which is crucial for the performance of anomaly detection models. The following steps were applied: 
\begin{enumerate}
    \item \textbf{Header filters:} Applied to control which information is included in HTTP headers during analysis. This helps eliminate redundant or noisy data, improving the relevance of the analyzed information.
    \item \textbf{urlDecode:} Next, decoding the input to prevent it from being URL-encoded (e.g., converting "\%20" to blank spaces) to handle data from URLs. This ensures that information is correctly interpreted and prevents errors due to malformed data.
    \item \textbf{decode('utf-8'):} Then, converting UTF-8 encoded byte sequences into Unicode strings. This is crucial for handling data containing special or international characters, ensuring subsequent analysis can properly process this data.
    \item \textbf{urlDecode (second time):} Similar to step 2, decoding the input again in URL format to correctly interpret it and avoid misinterpretation or malicious data manipulation errors from URLs. Attackers often use double encoding to mask attacks in the URL.
    \item \textbf{lowercase:} Finally, converting all input characters to lowercase to normalize the data. This facilitates comparison and feature search regardless of whether characters are uppercase or lowercase, helping to avoid case sensitivity issues.
\end{enumerate}
These steps adequately prepare the data for subsequent analysis stages, improving the accuracy and efficiency of anomaly detection models in web applications.

\subsection{Dictionary Creation}
The algorithm for constructing a dictionary of terms, described in Algorithm \ref{algorithm:dictionary},  uses CountVectorizer to create a set of tokens from generic attack datasets and valid application datasets. The process is divided into three main steps: data preprocessing, tokenization, and dictionary construction.

\begin{algorithm}[h!]
\caption{Dictionary Construction with CountVectorizer}
\begin{algorithmic}[1]
    \REQUIRE \textit{Generic\_Attacks\_Data, Normal\_App\_Data} \\ Generic attack datasets and Normal application dataset.
    \ENSURE Set of tokens
    \STATE \textbf{Step 1: Data Preprocessing}
    \FORALL{request in input data}
        \STATE Apply header filters
        \STATE Decode using urlDecode
        \STATE Decode to Unicode using decode('utf-8')
        \STATE Decode using urlDecode
        \STATE Convert request to lowercase
    \ENDFOR
    \STATE \textbf{Step 2: Tokenization}
    \STATE Use CountVectorizer to tokenize preprocessed requests.
    \STATE \textbf{Step 3: Dictionary Construction}
    \STATE Build a set of unique terms found.
    \RETURN Sets of terms
\end{algorithmic}
\label{algorithm:dictionary}
\end{algorithm}

In data preprocessing, each request for input data undergoes several preprocessing steps to standardize and prepare the data for tokenization. In this step, the processes previously detailed in subsection \ref{subsection:preprocessingstage} are applied. After preprocessing, requests are tokenized using \texttt{CountVectorizer}. Bearing in mind that some attacks use specially crafted input with special characters (e.g., ., ;, <, >, =, /), we implement the use of spaces as a separator as defined in \cite{rmartinez:thesis}. This process breaks requests into individual tokens (terms), and eventually a set of unique terms (tokens) is created from the tokenized requests. This set serves as a dictionary used for subsequent analysis.

The goal of creating this dictionary is to ensure that tokens that may be present in attacks, regardless of the specific application, are considered during subsequent feature selection.

To validate our approach with a real-life application, we use the DRUPAL dataset created in \cite{rmartinez:thesis}, based on real traffic to the public website of a service of a Uruguayan university. This dataset was generated by logging website traffic with ModSecurity for three full days, resulting in a total of 65582 valid entries. This dataset was used to build the dictionary and later in the training of the model.

\subsection{Applying Mutual Information}  Mutual Information is a measure used in data analysis to assess the dependence between two random variables. It helps in selecting relevant features. In the context of web attack detection, it is used to identify which features or tokens are more informative in differentiating between normal requests and potentially malicious ones.

\begin{algorithm}[h!]
\caption{Mutual Information with TF-IDF Vectorization}
\begin{algorithmic}[1]
    \REQUIRE Dataset $\mathbf{D}$, Set of web requests $R = \{r_1, r_2, \ldots, r_n\}$
    \ENSURE Set of tokens ordered by mutual information value $\mathbf{T}$
    \STATE $DIC \gets \text{dictionary initialization}$ \COMMENT{$DIC = \{f_1, f_2, \ldots, f_n\}$}
    \STATE $X \gets D_{\neg\text{target}}$ 
    \STATE $Y \gets D_{\text{target}}$  \COMMENT{ $Y$ is target vector $\{t_i\}$}
    \STATE $TFIDF \gets \text{TfidfVectorizer(vocabulary=DIC)}$
    \STATE $X_{tfidf} \gets TFIDF.\text{fit\_transform}(X)$ \COMMENT{ where $X_{tfidf}$ is the feature matrix $\{r_i \times f_i\}$, applying TF-IDF vectorization to all requests in $R$}
    \STATE $ICM \gets \text{mutual\_info\_classif}(X_{tfidf}, Y)$
    \FOR{$f_i \in DIC$}
        \STATE $\mathbf{F_i} \gets (f_i, ICM_i)$
    \ENDFOR
    \STATE $\mathbf{F} \gets \text{sorted}(\mathbf{F})$
    \RETURN $\mathbf{F}$
\end{algorithmic}
\label{algorithm:mutualInformation}
\end{algorithm}

The feature selection process, described in Algorithm \ref{algorithm:mutualInformation}, utilizes the set of valid web requests from the application and the set of attacks defined in Subsection \ref{subsubsection:attackdatasets}. The union of these sets is defined as $R = \{r_1, r_2, \ldots, r_n\}$.The dictionary is then constructed using Algorithm \ref{algorithm:dictionary}. Subsequently, all requests are vectorized using TF-IDF, generating the matrix $X_{tfidf}$. Mutual information values between each feature and the target are calculated for each feature in the dictionary, using $X_{tfidf}$ and the set of labels $Y$. Finally, the ordered set of feature, based on mutual information, $\mathbf{F}$ is returned.

Once the characteristics are ordered according to the corresponding mutual information value, Algorithm \ref{algorithm:anomalyDetection} is applied, which uses a bag-of-words model that tokenizes each request \( r_i \) in a sequence \( r_i = \{token_{i1}, \ldots, token_{in_i}\} \), where \( n_i \) denotes the number of tokens selected in the request. These representations transform the set of normal HTTP requests into input vectors for use in the one-class classification (OCC) model using a one-class support vector machine (OCSVM). This approach assumes a scenario where only valid requests are available, without the need for labeled attack samples. The OCSVM is trained using these representations, with the goal of distinguishing normal traffic from possible anomalies. By mapping data to the feature space defined by the kernel, OCSVM identifies inliers (normal requests) and outliers (potential attacks) based on their distance from a separating hyperplane.

\begin{algorithm}[h!]
\caption{Anomaly Detection with BoW and One-Class SVM}
\begin{algorithmic}[1]
\STATE \textbf{Input:} Set of normal HTTP requests $D = \{r_1, r_2, \ldots, r_m\}$
\STATE \textbf{Output:} One-Class SVM Model $model$
\STATE \textbf{Step 1: Preprocessing}
\FOR{each request $r_i \in D$}
    \STATE Apply header filters, URL decoding, and Unicode conversion to $r_i$
    \STATE Convert $r_i$ to lowercase
\ENDFOR

\STATE \textbf{Step 2: Feature Selection and Tokenization}
\STATE Select $\mathbf{N}$ relevant tokens $\mathbf{T}$ based on Algorithm \ref{algorithm:mutualInformation}
\STATE Tokenize $r_i$ into tokens $r_i = \{token_{i1}, \ldots, token_{i_N}\}$

\STATE \textbf{Step 3: Vectorization using Bag of Words (BoW)}
\STATE Create BoW representations for each request:
\FOR{each request $r_i \in D$}
    \STATE Construct a BoW vector $\mathbf{v}_i$ using selected tokens $\mathbf{T}$
\ENDFOR

\STATE \textbf{Step 4: One-Class SVM Training}
\STATE Train One-Class SVM model using BoW vectors $\{\mathbf{v}_1, \ldots, \mathbf{v}_m\}$

\STATE \textbf{Step 5: Parameter Optimization}
\STATE Perform grid search to optimize parameters $ \nu $ and $ \gamma $ for the OCSVM model

\STATE \textbf{Output:} Trained One-Class SVM Model $model$
\end{algorithmic}
\label{algorithm:anomalyDetection}
\end{algorithm}

The optimal operational threshold \( \theta \) is determined using a grid search approach, varying \( \theta \) across a range and evaluating its impact on the Receiver Operating Characteristic (ROC) curve to achieve optimal performance metrics. Specifically, parameter optimization involves selecting \( \nu \) and \( \gamma \), where \( \gamma \) defines the kernel's frontier and \( \nu \) represents the probability of encountering a new, but normal, observation outside this frontier. The grid search method employs a modified performance metric \( \hat{F} \), akin to the F1-score, derived from normal and unlabeled examples, to determine the best parameters \( \nu = 0.05 \) and \( \gamma = 0.5 \) for the OCSVM classifier, ensuring robust anomaly detection in web application security contexts \cite{montes2021web}.

\section{Discussion}
\label{sec:discuss}
This section discusses the outcomes of our experiment, which was designed to assess the effectiveness of the proposed feature selection algorithm in identifying web attacks.

As an initial stage of the experiment, we combined:
 \begin{enumerate}
     \item Valid data from the Drupal application with the set of varied attacks (in this case built with SR-BH 2020)
     \item Valid data from the SR-BH 2020 application with the set of varied attacks (in this case built using PKDD)
 \end{enumerate}
 
Using Algorithm \ref{algorithm:mutualInformation}, feature sets of sizes 50, 100, 150, and 200 were selected to evaluate the effectiveness of training the model with different dimensions. The experiment was conducted on both sets to assess the adaptability of the approach to different attack scenarios and traffic types.

\begin{figure*}[ht!]
    \centering
    \begin{minipage}{0.45\textwidth}
        \centering
        \includegraphics[width=\textwidth]{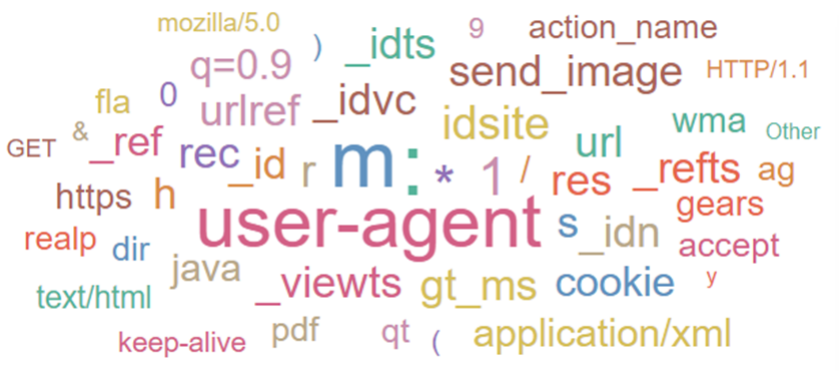}
        \caption{Top 50 Feature Selection Drupal}
        \label{fig:top50Drupal}
    \end{minipage}\hfill
    \begin{minipage}{0.45\textwidth}
        \centering
        \includegraphics[width=\textwidth]{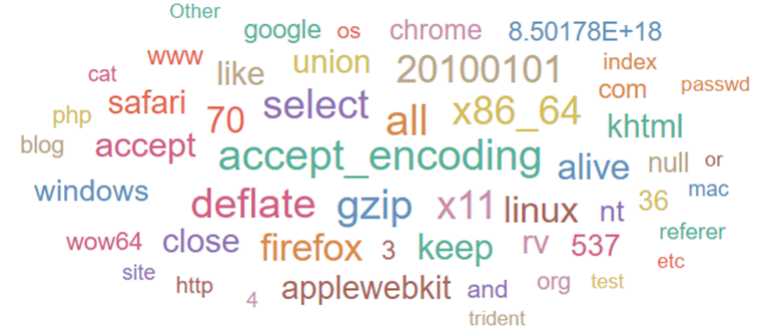}
        \caption{Top 50 Feature Selection SR-BH 2020}
        \label{fig:top50Capec}
    \end{minipage}
\end{figure*}

Figures \ref{fig:top50Drupal} and \ref{fig:top50Capec} show the top 50 features selected for the Drupal and SR-BH 2020 datasets, respectively.

In a second stage, Algorithm \ref{algorithm:anomalyDetection} was used to train the One-Class SVM models using the feature sets derived from the previous stage. Each model was evaluated with a test set containing valid data and application-specific attacks to assess its ability to distinguish between normal traffic and potential attacks.

Although a set of attacks (generic attacks) is available, a multi-class supervised model was not chosen. This decision is made after training a Random Forest with n\_estimators=100 and random\_state=42. This classifier produced a good performance in training (FPR: 0.00002, TPR: 1.00) but when evaluated with validation data of the application to be protected, the accuracy decreased significantly (FPR: 0.0161, TPR: 0.22) for the attack class. This indicates that the model cannot generalize to new data and is ineffective in real environments.

The test datasets were structured as follows:
\begin{itemize}
    \item \textbf{Drupal Case:} Containing 19,679 valid data and 382 incidents identified as attack
    \item \textbf{SR-BH 2020 Case:} Containing 52,520  valid data and 38,262 incidents identified as attack
\end{itemize}

The performance of the proposed method is analyzed in terms of True Positive Rate (TPR) and False Positive Rate (FPR). In our case, TPR and FPR indicate the ratio of requests correctly and incorrectly classified as attacks, respectively.

Table \ref{tab:result} presents performance metrics for the Drupal case. Increasing the features from 50 to 100 significantly improves the TPR to 91.76\%, maintaining a balanced FPR of 2.29\% and a high AUC of 0.97. Beyond 100 features, the TPR decreases and the FPR varies, indicating the 100-feature set selected by the algorithm is the most effective.

Compared to the expert-assisted method, which achieves the highest accuracy of 98.4\%, it shows a TPR of 81.36\% and an FPR of 1.23\%. While accurate, this method demonstrates a trade-off in TPR compared to the 100-feature set, highlighting the benefit of custom feature selection in attack detection scenarios.

\begin{figure}[h!]
    \centering
    \includegraphics[width=1\linewidth]{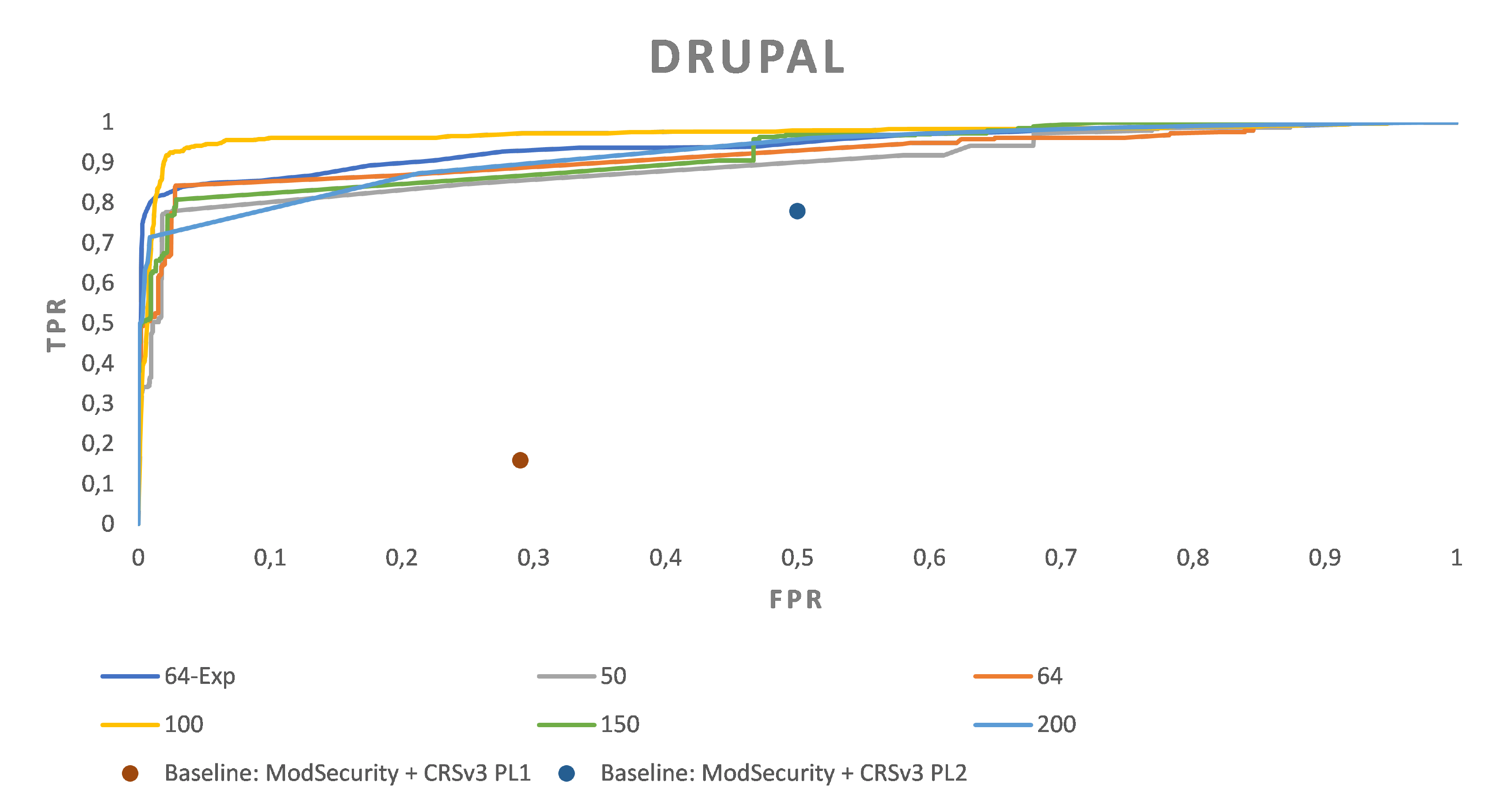}
    \caption{ROC curve - Drupal dataset}
    \label{fig:curvaRoc Drupal}
\end{figure}

\begin{table}[h!]
    \centering
     \begin{tabular}{|l|l|l|l|l|}
     \hline
        \textbf{N} & \textbf{ACC} & \textbf{TPR (\%)} & \textbf{FPR(\%)} & \textbf{AUC} \\ 
     \hline
        Expert assisted  & 0.984 & 81.36 & 1.23 & 0.93\\
        50 & 0.977 & 76.99 & 1.83 & 0.90\\
        64 & 0.968 & 84.75 & 2.80 & 0.92\\ 
        100 & 0.978 & 91.76 & 2.29 & 0.97\\
        150 & 0.973 & 76.75 & 2.24 & 0.92\\
        200 & 0.714 & 71.42 & 0.90 & 0.92\\
    \hline
    \end{tabular}
    \caption{Performance Results Drupal}
    \label{tab:result}
\end{table}
Figure \ref{fig:curvaRoc Drupal} shows the ROC curves for different feature sets (50, 64, 100, 150, 200 tokens) and reference comparisons (ModSecurity with CRS at PL1 and PL2 levels). Higher AUC values indicate better performance, and curves closer to the upper-left corner represent more accurate models.

Comparing these ROC curves, we identify that the optimal feature set for distinguishing between normal and malicious web requests is the set of 100 tokens. This set is compared to a set of 64 tokens selected by an expert, shown in the blue curve, which slightly improves the FPR but decreases the TPR across all other curves of 64, 100, and 150 tokens.

The points on the curve represent the performance of ModSecurity, showing that there are several points where the 100-feature model outperforms both ModSecurity configurations. When comparing the results with the baseline, the best ModSecurity configuration detects 75\% of the attacks, while the model with 100 features detects 91.76\%. In terms of FPR, ModSecurity has a rate of 39.69\%, whereas the model achieves a significantly lower rate of 2.29\%. Therefore, the 100-feature model outperforms ModSecurity.

When comparing the model based on 100 selected features with results from previous studies, such as \cite{rmartinez:thesis} achieving a TPR of 94.43\% and an FPR of 6.00\%, and the RoBERTa + OCSVM \cite{montes2021web} model with a slightly higher TPR of 95.00\% and a higher FPR low of 3.73\%, it is observed that the 100-feature model achieves a comparable TPR of 91.76\% with a significantly lower FPR of 2.29\%. This demonstrates good performance in maintaining a high attack detection rate while minimizing false positives achieved with low computational cost and without expert intervention. Furthermore, this was achieved using a standard one-class classifier, showing that classic techniques can produce good performance in this problem when the correct set of features is selected.  

Table \ref{tab:resultSR-BH 2020} compares the performance results using different numbers of features in detecting attacks on the SR-BH 2020 dataset, as in the previous case by increasing the number of features from 50 to 100 significantly improves the TPR, increasing from 66.92\% to 78.87\% while maintaining a reasonable FPR of 5.18\% and a high AUC value of 0.84, demonstrating a good discrimination capacity between normal and attack traffic.

However, when increasing beyond 100 features, it is observed that the 150 feature set achieves a TPR of 71.87\% with a low FPR of 3.36\%, and the 200 feature set shows a TPR of 73.54\% with an FPR of 6.80\%. , both have a comparable AUC of 0.81 but lower than that of 100 features.

\begin{table}[h!]
    \centering
     \begin{tabular}{|l|l|l|l|l|}
     \hline
        \textbf{N} & \textbf{ACC} & \textbf{TPR (\%)} & \textbf{FPR(\%)} & \textbf{AUC} \\ 
     \hline
        Expert assisted & 0.783 & 51.39 & 2.08 & 0.72\\
        50 & 0.839 & 66.92 & 3.70 & 0.70\\
        64 & 0.843 & 68.20 & 3.92 & 0.67\\ 
        100 & 0.881 & 78.87 & 5.18 & 0.84\\
        150 &  0.861 & 71.87 & 3.36& 0.81\\
        200 & 0.849 & 73.54 & 6.80 & 0.81\\
    \hline
    \end{tabular}
    \caption{Performance Results SR-BH 2020
}
    \label{tab:resultSR-BH 2020}
\end{table}

\begin{figure} [h!]
    \centering
    \includegraphics[width=1\linewidth]{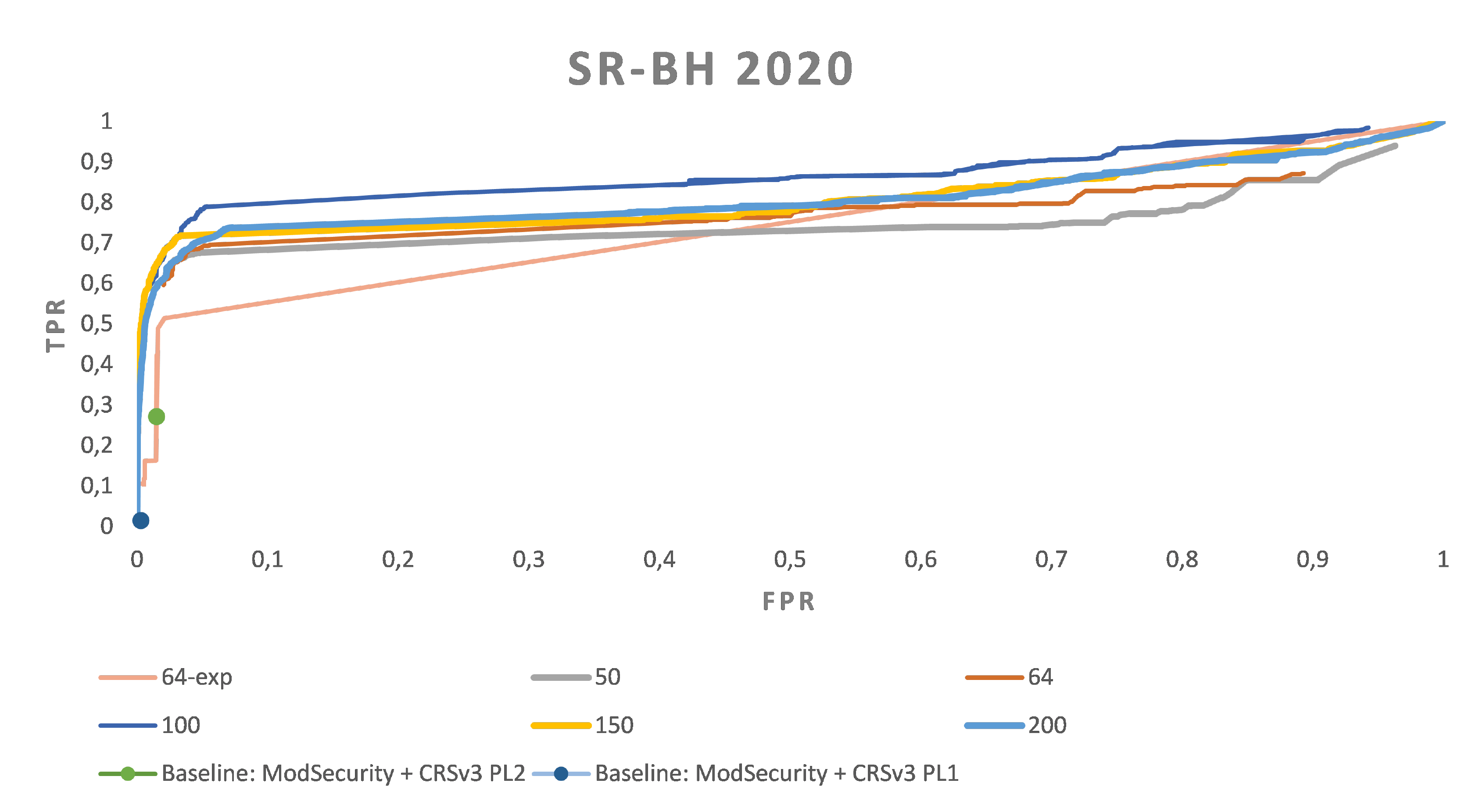}
    \caption{ROC curve - SR-BH 2020 dataset}
    \label{fig:CurvasROCSR-BH 2020}
\end{figure}

Figure \ref{fig:CurvasROCSR-BH 2020} shows the ROC curves for the SR-BH 2020 case. Feature sets of 100 and 150 dimensions offer the best performance, balancing high TPR and low FPR, and outperforming the ModSecurity baselines. The 64-exp model, with 64 features selected by experts, underperforms compared to higher-dimension sets. While the ModSecurity baselines provide a reasonable FPR (1,47\%), their TPR(27,10\%) is very low, indicating that custom feature selection significantly improves precision.
The experiments showed that the new feature selection algorithm performs better than the expert-assisted version in terms of performance, while also decreasing the need for human involvement in the feature selection process. This indicates that the new feature selection method can enhance effective attack detection, particularly with feature sets containing 100-150 dimensions.

Results from models using 100-150 dimensions could suggest that using wrapper methods for feature selection could potentially improve performance in detecting web attacks by more effectively prioritizing relevant features and achieving a better balance between the TPR and FPR values.

\section{Related Work}
\label{sec:relwork}

The research discussed in \cite{betarte2018web, rmartinez:thesis} uses machine learning and pattern recognition methodologies to address false positives. The study presents four different approaches to detect web application attacks in various scenarios. It employs the Bag-of-Words model for feature extraction, and feature selection is carried out by experts. The study further evaluates the performance of several supervised classification models in comparison to each other and a reference ModSecurity web application firewall (WAF).

The application of deep learning techniques for the detection and prevention of attacks on web applications is explored in the study reported in \cite{montes2021web}. Unlike prior research, this work frames the problem as a supervised classification task and employs a pre-trained RoBERTa model to the dataset. This pre-trained model is utilized to extract features from HTTP requests, which are subsequently employed to train a one-class classifier. 

While the approaches presented in \cite{betarte2018web, rmartinez:thesis,montes2021web} have shown promising results in terms of TPR/FPR, they both also have weaknesses. The approach in \cite{betarte2018web, rmartinez:thesis} is distinguished by its computational efficiency and TPR/FPR rate, although it has limitations such as the need for labeled data to train the model and a security expert to define the set of features to be used. On the other hand, the method presented in \cite{montes2021web} has slightly better results without requiring the intervention of a security expert considering contextual information. However, this approach has the limitation of requiring training the RoBERTa model for each dataset, which is computationally expensive, and classification time increases significantly.

These limitations give rise to two distinct challenges. The first challenge is to devise a pipeline design that makes use of RoBERTa and enables the utilization of a RoBERTa model trained on one dataset for a different one. One potential solution could involve reducing dimensionality and disregarding certain context features, instead focusing solely on the important ones. However, this gives rise to the second challenge, which also existed in more traditional approaches: the necessity for an expert to select relevant features.

In the work \cite{marcilio2020explanations}, SHAP is analyzed as a feature selection mechanism. SHAP is a model-agnostic approach that assigns feature importance based on their contribution to the model's outcome. The proposal uses these contributions to rank features according to their importance as a feature selection strategy, demonstrating superiority over other more common mechanisms.

The comparative study conducted in \cite{doi:10.1177/0165551518770967} compares mutual information  with sensitivity analysis (DSA) for feature selection in a banking telemarketing data set concluding that mutual information efficiently identifies relevant features, which reduces the redundancy and enables faster and more accurate customer subscription modeling. It highlights the model's ability to handle large data sets by quickly evaluating the information content. It establishes that despite MI's older methodology, it competes favorably with newer approaches in this case DSA proving to be effective and making it a solid choice for applications that prioritize prediction accuracy over computational complexity.

The work presented in \cite{azmi2021feature} investigates the use of feature selection techniques to improve the detection of distributed denial of service (DDoS) attacks employing machine learning algorithms. Features are selected from the UNSW-NB 15 dataset\cite{7348942} using methods such as information gain and data reduction. Then, the selected features are classified using Artificial Neural Network (ANN), Naïve Bayes and Decision Table algorithms. Comparative analysis with previous studies using the same data set confirms the effectiveness of the feature selection approach, demonstrating higher accuracy of the classifier in detecting DDoS attacks.
\section{Conclusion and Further Work}
\label{sec:concl}
We have presented a method for training and evaluating an anomaly detection model using One-Class SVM. The approach involves using feature selection based on mutual information. The results indicate that a 100-dimensional feature set achieved the best balance between true positive rate (TPR) and false positive rate (FPR), outperforming the expert-assisted method. This demonstrates the effectiveness of feature selection based on mutual information in identifying the most relevant features and improving model performance. Even though the expert-assisted method achieved higher accuracy, it had a lower true positive rate, highlighting the importance of automated approaches in feature selection.

For future research, we aim to investigate feature selection using wrapper-based methods , such as Recursive Feature Elimination (RFE) and forward selection algorithms \cite{chandrashekar2014survey}. As observed in the second experiment of the preceding section \ref{sec:discuss}, these methods may positively impact performance by considering feature interactions with the learning algorithm, leading to more precise and tailored model selection. Although more computationally intensive, they can yield significant improvements in model performance compared to the current mutual information-based approach.

The tokenizer used in this research is a general natural language tokenizer. In future work, we intend to create a tokenizer tailored for the HTTP language, which will consider the context and structure of the language, potentially resulting in enhanced accuracy. This specialized tokenizer would more effectively capture the syntactic and semantic relationships between request features, consequently helping to improve the overall effectiveness of anomaly detection systems in web security.

\bibliographystyle{plain}
\bibliography{wafintl}

\begin{thebibliography}{10}

\bibitem{ecml-pkdd-challenge}
Analyzing web traffic: Ecml/pkdd 2007 discovery challenge.
\newblock \url{http://www.lirmm.fr/pkdd2007-challenge/}.

\bibitem{applebaum2021signature}
Simon Applebaum, Tarek Gaber, and Ali Ahmed.
\newblock Signature-based and machine-learning-based web application firewalls:
  A short survey.
\newblock {\em Procedia Computer Science}, 189:359--367, 2021.

\bibitem{azmi2021feature}
Muhammad Aqil~Haqeemi Azmi, Cik Feresa~Mohd Foozy, Khairul Amin~Mohamad Sukri,
  Nurul~Azma Abdullah, Isredza Rahmi~A Hamid, and Hidra Amnur.
\newblock Feature selection approach to detect ddos attack using machine
  learning algorithms.
\newblock {\em JOIV: International Journal on Informatics Visualization},
  5(4):395--401, 2021.

\bibitem{doi:10.1177/0165551518770967}
Néstor Barraza, Sérgio Moro, Marcelo Ferreyra, and Adolfo de~la Peña.
\newblock Mutual information and sensitivity analysis for feature selection in
  customer targeting: A comparative study.
\newblock {\em Journal of Information Science}, 45(1):53--67, 2019.

\bibitem{beraha2019feature}
Mario Beraha, Alberto~Maria Metelli, Matteo Papini, Andrea Tirinzoni, and
  Marcello Restelli.
\newblock Feature selection via mutual information: New theoretical insights.
\newblock In {\em 2019 international joint conference on neural networks
  (IJCNN)}, pages 1--9. IEEE, 2019.

\bibitem{betarte2018improving}
Gustavo Betarte, Eduardo Giménez, Rodrigo Martinez, and {\'A}lvaro Pardo.
\newblock Improving web application firewalls through anomaly detection.
\newblock In {\em 2018 17th IEEE International Conference on Machine Learning
  and Applications (ICMLA)}, pages 779--784. IEEE, 2018.

\bibitem{betarte2018web}
Gustavo Betarte, Rodrigo Mart{\'\i}nez, and {\'A}lvaro Pardo.
\newblock Web application attacks detection using machine learning techniques.
\newblock In {\em 2018 17th IEEE International Conference on Machine Learning
  and Applications (ICMLA)}, pages 1065--1072. IEEE, 2018.

\bibitem{breiman1984classification}
Leo Breiman, Jerome Friedman, Richard Olshen, and Charles Stone.
\newblock {\em Classification and regression trees}.
\newblock Wadsworth International Group, 1984.

\bibitem{chandrashekar2014survey}
Girish Chandrashekar and Ferat Sahin.
\newblock A survey on feature selection methods.
\newblock {\em Computers \& electrical engineering}, 40(1):16--28, 2014.

\bibitem{chaudhari2014survey}
Gopal~R Chaudhari and Madhav~V Vaidya.
\newblock A survey on security and vulnerabilities of web application.
\newblock {\em International Journal of Computer Science and Information
  Technologies}, 5(2):1856--1860, 2014.

\bibitem{cover1991elements}
Thomas~M Cover and Joy~A Thomas.
\newblock {\em Elements of Information Theory}.
\newblock John Wiley \& Sons, 1991.

\bibitem{devlin2018bert}
Jacob Devlin, Ming-Wei Chang, Kenton Lee, and Kristina Toutanova.
\newblock Bert: Pre-training of deep bidirectional transformers for language
  understanding.
\newblock {\em arXiv preprint arXiv:1810.04805}, 2018.

\bibitem{gallagher2009classification}
Brian Gallagher and Tina Eliassi-Rad.
\newblock Classification of http attacks: a study on the ecml/pkdd 2007
  discovery challenge.
\newblock Technical report, Lawrence Livermore National Laboratory (LLNL),
  Livermore, CA, July 2009.

\bibitem{ghanbari2015comparative}
Z~Ghanbari, Y~Rahmani, H~Ghaffarian, and M~Hossein Ahmadzadegan.
\newblock Comparative approach to web application firewalls.
\newblock In {\em 2015 2nd International Conference on Knowledge-Based
  Engineering and Innovation (KBEI)}, pages 808--812. IEEE, 2015.

\bibitem{gniewkowski2023sec2vec}
Mateusz Gniewkowski, Henryk Maciejewski, Tomasz Surmacz, and Wiktor
  Walentynowicz.
\newblock Sec2vec: Anomaly detection in http traffic and malicious urls.
\newblock In {\em Proceedings of the 38th ACM/SIGAPP Symposium on Applied
  Computing}, pages 1154--1162, 2023.

\bibitem{guyon2002gene}
Isabelle Guyon, Jason Weston, Stephen Barnhill, and Vladimir Vapnik.
\newblock Gene selection for cancer classification using support vector
  machines.
\newblock {\em Machine learning}, 46(1):389--422, 2002.

\bibitem{kohavi1997wrappers}
Ron Kohavi and George~H John.
\newblock Wrappers for feature subset selection.
\newblock In {\em Readings in machine learning}, pages 273--324. Elsevier,
  1997.

\bibitem{li2020weighted}
Jieling Li, Hao Zhang, and Zhiqiang Wei.
\newblock The weighted word2vec paragraph vectors for anomaly detection over
  http traffic.
\newblock {\em IEEE Access}, 8:141787--141798, 2020.

\bibitem{liu2021http}
Bocheng Liu and Fan Yang.
\newblock Http traffic analysis based on multiple deep convolution network
  model generation algorithms.
\newblock In {\em Journal of Physics: Conference Series}, volume 2037, page
  012048. IOP Publishing, 2021.

\bibitem{liu2019roberta}
Yinhan Liu, Myle Ott, Naman Goyal, Jingfei Du, Mandar Joshi, Danqi Chen, Omer
  Levy, Mike Lewis, Luke Zettlemoyer, and Veselin Stoyanov.
\newblock Roberta: A robustly optimized bert pretraining approach.
\newblock {\em arXiv preprint arXiv:1907.11692}, 2019.

\bibitem{marcilio2020explanations}
Wilson~E Marc{\'\i}lio and Danilo~M Eler.
\newblock From explanations to feature selection: assessing shap values as
  feature selection mechanism.
\newblock In {\em 2020 33rd SIBGRAPI conference on Graphics, Patterns and
  Images (SIBGRAPI)}, pages 340--347. Ieee, 2020.

\bibitem{rmartinez-ladc-2018}
Rodrigo Martinez.
\newblock Enhancing web application attack detection using machine learning.
\newblock {\em 8th Latin-American Symposium on Dependable Computing}, 2018.

\bibitem{rmartinez:thesis}
Rodrigo Martínez.
\newblock {Enhancing web application attack detection using machine learning}.
\newblock Master's thesis, Facultad de Ingeniería,UdelaR - Área Informática
  del Pedeciba, Uruguay, 2019.

\bibitem{miao2016survey}
Jianyu Miao and Lingfeng Niu.
\newblock A survey on feature selection.
\newblock {\em Procedia computer science}, 91:919--926, 2016.

\bibitem{montaruli2023adversarial}
Biagio Montaruli, Luca Demetrio, Andrea Valenza, Battista Biggio, Luca
  Compagna, Davide Balzarotti, Davide Ariu, and Luca Piras.
\newblock Adversarial modsecurity: Countering adversarial sql injections with
  robust machine learning.
\newblock {\em arXiv preprint arXiv:2308.04964}, 2023.

\bibitem{montes2021web}
Nicol{\'a}s Montes, Gustavo Betarte, Rodrigo Mart{\'\i}nez, and Alvaro Pardo.
\newblock Web application attacks detection using deep learning.
\newblock In {\em Progress in Pattern Recognition, Image Analysis, Computer
  Vision, and Applications: 25th Iberoamerican Congress, CIARP 2021, Porto,
  Portugal, May 10--13, 2021, Revised Selected Papers 25}, pages 227--236.
  Springer, 2021.

\bibitem{7348942}
Nour Moustafa and Jill Slay.
\newblock Unsw-nb15: a comprehensive data set for network intrusion detection
  systems (unsw-nb15 network data set).
\newblock In {\em 2015 Military Communications and Information Systems
  Conference (MilCIS)}, pages 1--6, 2015.

\bibitem{owasp}
{OWASP}.
\newblock Open web application security project.
\newblock https://www.owasp.org.

\bibitem{ren2018web}
Xin Ren, Yupeng Hu, Wenxin Kuang, and Mohamadou~Ballo Souleymanou.
\newblock A web attack detection technology based on bag of words and hidden
  markov model.
\newblock In {\em 2018 IEEE 15th International Conference on Mobile Ad Hoc and
  Sensor Systems (MASS)}, pages 526--531. IEEE, 2018.

\bibitem{riera2022new}
Tom{\'a}s~Sureda Riera, Juan-Ram{\'o}n~Bermejo Higuera, Javier~Bermejo Higuera,
  Jos{\'e}-Javier~Mart{\'\i}nez Herraiz, and Juan-Antonio~Sicilia Montalvo.
\newblock A new multi-label dataset for web attacks capec classification using
  machine learning techniques.
\newblock {\em Computers \& Security}, 120:102788, 2022.

\bibitem{romanartificial}
Jes{\'u}s-{\'A}ngel Rom{\'a}n-Gallego, Mar{\'\i}a-Luisa P{\'e}rez-Delgado,
  Marcos~Luengo Vi{\~n}uela, and Mar{\'\i}a-Concepci{\'o}n Vega-Hern{\'a}ndez.
\newblock Artificial intelligence web application firewall for advanced
  detection of web injection attacks.
\newblock {\em Expert Systems}, page e13505.

\bibitem{salehin2024automl}
Imrus Salehin, Md~Shamiul Islam, Pritom Saha, SM~Noman, Azra Tuni, Md~Mehedi
  Hasan, and Md~Abu Baten.
\newblock Automl: A systematic review on automated machine learning with neural
  architecture search.
\newblock {\em Journal of Information and Intelligence}, 2(1):52--81, 2024.

\bibitem{tibshirani1996regression}
Robert Tibshirani.
\newblock Regression shrinkage and selection via the lasso.
\newblock {\em Journal of the Royal Statistical Society: Series B
  (Methodological)}, 58(1):267--288, 1996.

\end{thebibliography}

\end{document}